
\magnification=\magstep1
\baselineskip=24 true pt
\hsize=34 pc
\vsize=45 pc
\bigskip
\bigskip
\rightline {IP/BBSR/92-77}
\rightline {November, 1992}
\centerline {\bf WORMHOLES AND SYMMETRIES OF STRING EFFECTIVE ACTION}
\vfil
\centerline {{\bf S. Pratik Khastgir and Jnanadeva
Maharana}\footnote\dag{e-mail: pratik/maharana@iopb.ernet.in
}}
\bigskip
\centerline {\it { Institute of Physics, Bhubaneswar-751005, INDIA}}
\vfil
\centerline {\bf Abstract}

The dimensional reduction technique is adopted to derive string
effective action. Wormhole solutions corresponding to space-time
geometries $R^1\times S^1\times S^2$ and $R^1\times S^3$ are
presented. The duality and $SL(2,R)$ symmetries are implemented to
generate new wormhole solutions.
\vfil
\eject

 The topology changing processes play a very important role in
quantum gravity and their effects are surprising [1].
 It has been argued that such processes give rise to
loss of quantum coherence [2]. On the other hand, Coleman has advocated
that the the wormholes introduce quantum indeterminacy [3] in constants
of Nature. Thus it has been proposed that the fundamental constants
are endowed with a
probability distribution function for universes and it is necessary
to take ensemble average over all the universes while computing them [4].
The consequences of topology changing processes have attracted
considerable attention in the recent past and have been explored extensively
[5]. If the wormhole corresponds to a saddle point of the Euclidean
action, then the semiclassical approximations can be employed in
order to facilitate computations of vertex operators and correlation
functions in the path integral formalism [6].
Indeed, axionic wormhole solutions have been found by Giddings and
Strominger [1] in four dimensions
for effective actions that arise in string theories. These
actions are to be envisaged as derived from tree level effective
action of bosonic/heterotic string theories in critical dimensions
where the internal dimensions are compactified on suitable geometries.

Recently Schwarz and one of us [7] have employed the Scherk-Schwarz [8]
dimensional
reduction technique in order to derive string effective action in
lower dimensions and to investigate the origin of the noncompact
symmetries in string theory. The approach generalises the results of
Veneziano and collaborators [9] where the backgrounds are allowed to
carry time dependence only.

The purpose of this note is to employ the techniques of Ref. [7] to
construct four dimensional string effective action and seek for
wormhole solutions. Indeed, we present the existence of several
wormhole solutions and show how new wormhole solutions can be generated by
implementing the rich symmetry structure of the reduced action,
demonstrating the elegance and power of the technique.

We outline below the main results of this investigation. First we
present a wormhole solution for the geometry $R^1\times S^1\times S^2$.
When the coordinate corresponding to the $S^1$ is taken to be
compact, then the wormhole solution is analogous to the 2+1
dimensional magnetic wormhole. Next recall the known
wormhole solution [1] in our formalism giving the explicit form of the
background fields. The target space duality trnsformation on a
wormhole, characterized by the axionic charge, $Q$, generates another
wormhole with a different charge. We would like to mention that
besides duality there is an $SL(2,R)$ symmetry of the 4-dimensional effective
action and the $SL(2,R)$ transformation generates wormhole with
different axionic charge.

Let us recall the main results of Ref. [7] and set the notations. The
bosonic part of the effective action of the heterotic string in
 $\hat D=D+d$ Euclidean dimensions ($\hat D=10$ for critical case) is,

$$S_{\hat g} = \int d^{\hat D}x~ \sqrt{ \hat g}~ e^{-\hat\phi}
\big [-\hat R
(\hat g) - \hat g^{\hat \mu \hat \nu} \partial_{\hat \mu} \hat\phi
\partial_{\hat \nu} \hat\phi + {1 \over 12} ~ \hat H_{\hat \mu \hat \nu
\hat \rho} ~
\hat H^{\hat \mu \hat \nu \hat \rho}\big ].\eqno (1)$$

\noindent $\hat H$ is the field strength of antisymmetric tensor and $\hat
\phi$ is the dilaton. Here we have set all the nonabelian gauge field
backgrounds to zero. When the backgrounds are independent of the `internal'
coordinates $y^{\alpha}, \alpha=1,2..d$ and the internal space is
taken to be torus, the metric $\hat g_{\hat \mu \hat \nu} $
 can be decomposed as

$$\hat g_{\hat \mu \hat \nu} = \left (\matrix {g_{\mu \nu} +
A^{(1)\gamma}_{\mu} A^{(1)}_{\nu \gamma} &  A^{(1)}_{\mu \beta}\cr
A^{(1)}_{\nu \alpha} & G_{\alpha \beta}\cr}\right ),\eqno (2)$$

\noindent where $G_{\alpha \beta}$ is the internal metric and $g_{\mu\nu}$,
the $D$-dimensional space-time metric, depend on the coordinates $x^{\mu}$.
The dimensionally reduced action is,

$$\eqalign {S_{\hat g} =& \int d^Dx \sqrt {g}~ e^{-\phi}
\bigg\{- R - g^{\mu \nu}
\partial_{\mu} \phi \partial_{\nu} \phi +{1\over 12}H_{\mu \nu \rho} ~
H^{\mu \nu \rho}\cr
&- {1 \over 8} {\rm tr} (\partial_\mu M^{-1} \partial^\mu
M)+ {1 \over 4}
{\cal F}^i_{\mu \nu} (M^{-1})_{ij} {\cal F}^{\mu \nu j} \bigg\}}. \eqno (3)$$

\noindent Here $\phi=\hat\phi-{1\over 2}\log\det G$ is the shifted dilaton.
$$H_{\mu \nu \rho} = \partial_\mu B_{\nu \rho} - {1 \over 2} {\cal A}^i_\mu
\eta_{ij} {\cal F}^j_{\nu \rho} + ({\rm cyc.~ perms.}),$$
\noindent ${\cal F}^i_{\mu \nu}$ is the $2d$-component vector of field
strengths
$${\cal F}^i_{\mu \nu} = \pmatrix {F^{(1) \alpha}_{\mu \nu}
\cr F^{(2)}_{\mu \nu \alpha}\cr} = \partial_\mu {\cal A}^i_\nu - \partial_\nu
{\cal A}^i_\mu \,\, ,\eqno (4)$$
\noindent $A^{(2)}_{\mu \alpha} = \hat B_{\mu \alpha} + B_{\alpha \beta}
A^{(1) \beta}_{\mu}$ (recall $B_{\alpha \beta}=\hat B_{\alpha \beta}$), and
the $2d\times 2d$ matrices are
$$M = \pmatrix {G^{-1} & -G^{-1} B\cr
BG^{-1} & G - BG^{-1} B\cr},\qquad \eta =  \pmatrix {0 & 1\cr 1 & 0\cr}
\, .\eqno (5)$$

\noindent The action (3) is invariant under a global $O(d,d)$ transformation

 $$M \rightarrow \Omega M \Omega^T, \qquad \Omega^T \eta \Omega = \eta, \qquad
{\cal A}_{\mu}^i \rightarrow \Omega^i{}_j {\cal A}^j_\mu, \qquad
\Omega \in O(d,d). \eqno (6)$$
\noindent Note that $M\in O(d,d)$ also and $M^T\eta M=\eta$. The
background equations of motion can be derived from (3). The classical
solutions of string effective action correspond to different string vacuua
and are given by solutions for $M$,$\cal F$ and $\phi$.

In what follows, we assume $D$=4 and $d$=6 so that $\hat D$=10
(critical dimensions) and derive wormhole solutions in four dimensions.

{$\bf R^1\times S^1\times S^2$} Wormholes: Let us choose the backgrounds
$G_{\alpha\beta},B_{\alpha\beta},\phi,$ and ${\cal A}^i_{\mu}$ to be constants.
The action (3) reduces to

$$S_4 = \int d^4x \sqrt {g}~\bigg\{- R+{1\over 12}H_{\mu \nu
\rho} ~H^{\mu \nu \rho}\bigg\}. \eqno (7)$$

We choose the following ansatz for the 4-dimensional metric and the
antisymmetric tensor field strength;

$$ds^2=dt^2+dr^2+R^2(r)d\Omega_2^2\qquad H_{t\theta\phi}=2R_0
\sin\theta~ dt\wedge d\theta\wedge d\phi. \eqno(8)$$

\noindent The coordinate $t$ has a range $[0,2\pi]$. The (rr)
Einstein equation becomes,
$$R'^2=1-{R_0^2\over R^2}.\eqno (9)$$

\noindent Here and everywhere prime denotes derivative with respect
to $r$. The solution $R^2=R^2_0+r^2$, has $R(r)\rightarrow \pm r$
for large $r$, corresponding to the two asymptotically flat
Euclidean regions,
and has throat size $R_0$. The other three Einstein equations  and the
$H$ equations are automatically satisfied for this ansatz. The volume
integral of $H$ is the axionic charge, $Q$, proportional to $R_0$.

Notice that if $t$ is identified as a compact coordinate the
$B_{t\phi}$ component has the interpretation of a gauge field such that
{\bf A}=(0,0,2$R_0$(1-cos$\theta$)). The
corresponding field strength, $F^2$=$8R_0^2/R^4$, is analogous to that
of a magnetic
monopole with above ansatz (8) for $H$. Thus we recover the 2+1
dimensional monopole solution of Gupta $et~ al.$ [10].

In Ref. [11], wormhole solutions with $R^1\times S^1\times S^2$
geometry was derived with matter content of antisymmetric and gauge
field. The monopole like solution presented here is different
from those of Ref. [11].

{$\bf R^1\times S^3$} Wormholes: This
is another four dimensional solution discovered by Giddings
and Strominger [1] . Effective action is same as (7),
but now the solution has different topology viz. $R^1\times S^3$.
The line element and the
$H$ are given by

$$ds^2=dr^2+R^2(r)d\Omega_3^2,\qquad H_{\theta\psi\phi}=2\sqrt {3}
R_0^2 \sin^2\theta \sin\psi~ d\theta\wedge d\psi
\wedge d\phi. \eqno(10)$$

 The (rr) metric equation becomes,
$$R'^2=1-{R_0^4\over R^4}.\eqno (11)$$
\noindent Again the solution  has $R^2(r)\rightarrow r^2$
for large $r$, corresponding to the two asymptotically flat
Euclidean regions with a throat size $R_0$.

Next, we choose the background $\phi$=const. and set $H_{\mu\nu\rho}$=0
and ${\cal A}^i_{\mu}
$=0.  The spherically symmetric
ansatz
is chosen for $G_{\alpha\beta}$ and $B_{\alpha\beta}$ and the
$6\times 6$ matrix is represented as three blocks of $2\times 2$
matrices of the following form.
$$G+B = \pmatrix {\Sigma_1 & 0 & 0\cr 0 & \Sigma_2 & 0\cr
0 & 0 & \Sigma_3\cr},\qquad \Sigma_j= \pmatrix {e^{\lambda_jD_j(r)} &
a_j(r)\cr -a_j(r) & e^{\lambda_jD_j(r)}\cr}, \eqno (12)$$
\noindent no summation over the repeated index $j$ is understood
above. The effective action (3)
takes the form

$$S_4 = \int d^4x \sqrt {g}~\bigg [- R+{1\over 2}g^{rr} \sum_{j=1}^3
\bigg \{\lambda_j^2
(\partial_rD_j)^2 +e^{-2\lambda_jD_j}(\partial_ra_j)^2\bigg \}\bigg ]
\eqno (13)$$

\noindent The equations of motion are as follows
$$G_{rr}=R_{rr}-{1\over 2}g_{rr}R={1\over 4}\sum_{j=1}^3\bigg \{\lambda_j^2
(\partial_rD_j)^2 +e^{-2\lambda_jD_j}(\partial_ra_j)^2 \bigg \},\eqno (14)$$

$$G_{ii}=R_{ii}-{1\over 2}g_{ii}R=-{1\over 4}g_{ii}g^{rr}\sum_{j=1}^3
\bigg \{\lambda_j^2 (\partial_rD_j)^2
+e^{-2\lambda_jD_j}(\partial_ra_j)^2\bigg \}
,\qquad i=\theta,\psi,\phi\eqno (15)$$

$$\partial_r({\sqrt g}g^{rr}e^{-2\lambda_j D_j}\partial_r a_j)=0,\eqno(16)$$
\noindent and
$$\partial_r({\sqrt g}g^{rr}\lambda_j\partial_r D_j)
+{\sqrt g}g^{rr}e^{-2\lambda_jD_j}(\partial_r a_j)^2=0.\eqno(17)$$

\noindent We mention in passing that equations (16) and (17) hold for
each $j$=1,2 and 3; moreover these equations represent the axionic charge
conservation and the dilaton evolution respectively. Whereas
eqns. (14) and (15) are Einstein equations. The choice,

$$ds^2=dr^2+R^2(r)d\Omega_3^2\eqno(18a)$$
$$e^{-2\lambda_jD_j}={Q_j^2\over R^4},\qquad a_j=\pm{i\over Q_j}R^2R'
\eqno(18b)$$
\noindent with the relation $R'^2=1-{R_0^4\over R^4}$, satisfies all
the equations of motion.
Where $R(r)$ is the scale factor and $R_0$ gives size of the wormhole
neck.

It is worthwhile to point out that the axionic background envisaged
above has some similarity with the axion solution adopted by Giddings
and Strominger [12]. However, our backgrounds arise from the
prescriptions of toroidal compactifications [7]. On the other hand,in
Ref. [12],
the antisymmetric tensor field, $B_{\alpha\beta}$,
associated with the internal dimensions,
are the fundamental K{\"a}hler forms of the internal dimensions and
their $x$ and $y$ dependence are decomposed in a specific manner.

Symmetries of the effective action: (i) The action (13) is invariant under
global $O(6,6)$ transformation for space-time dependent $G$ and $B$.
The manifestly $O(6,6)$ invariant action can be obtained by
expressing (13) in terms of $M$ as in
(3). The  duality
transformations $M\rightarrow\ \eta M \eta =M^{-1}$ is a special form
of the above global noncompact transformations. The new
backgrounds (duality transformed) ${\tilde a_j}$ and ${\tilde D_j}$ are
given by
$$e^{-2\lambda_j\tilde {D}_j}=(e^{\lambda_jD_j}+e^{-\lambda_jD_j}a_j^2)^2=
\bigg ({{R_0^4}\over {Q_j^2}}\bigg )^2e^{-2\lambda_jD_j},\eqno (19a)$$

$$\tilde {a}_j=-(e^{\lambda_jD_j}+e^{-\lambda_jD_j}a_j^2)^{-1}
e^{-\lambda_jD_j}a_j=-{{Q_j^2}\over {R_0^4}}a_j.\eqno (19b)$$

\noindent It is interesting to note that for $Q_j=R_0^2$ the axionic
charges of the original theory and the transformed theory are the
same. Thus such a value of $Q_j$ corresponds to a self dual theory.

(ii) Furthermore the same action (13) is invariant under the global
$SL(2,R)$ transformations.  The $SL(2,R)$ symmetry of the string
effective action in 4-dimensions has been recently discussed [13].
Let us define the matrix

$$S_j=\bigg (\matrix {{e^{\lambda_jD_j}+e^{-\lambda_jD_j}a_j^2} &
{e^{-\lambda_jD_j}a_j}\cr {e^{-\lambda_jD_j}a_j}
& {e^{-\lambda_jD_j}} \cr }\bigg ) \eqno(20)$$

\noindent The action (13) can be rewritten in terms of $S_j$s as

$$S_4 = \int d^4x \sqrt {g}~\bigg [- R-{1\over 4}g^{rr} \sum_{j=1}^3
\bigg \{{\rm tr}(\partial_rS_j\partial_rS^{-1}_j)\bigg \}\bigg] \eqno (21)$$

\noindent The action (21) is invariant
under the $SL(2,R)$ transformations $S\rightarrow\ {\tilde S}={\cal T}^{T}S
{\cal T}$
where ${\cal T}$ is the $SL(2,R)$ matrix. It should be understood
that there is an $SL(2,R)$ for each $S_j$.
The $SL(2,R)$ symmetry discussed in [13]arises by combining the
dilaton and the axion (dual to $H$ field) system and implementing a
transformation involving both the fields. In contrast, the $SL(2,R)$
symmetry presented here involves the fields $D_j$ and $a_j$, both
arising from the toroidal compactification.Conesequently we can generate new
backgrounds ${\tilde a}$ and ${\tilde D}$ through the implementation
of $SL(2,R)$ transformations. A special such transformation,

$${\cal T}=\bigg (\matrix {{0} &
{1}\cr {-1}
& {0} \cr }\bigg ), \eqno(22)$$

\noindent takes $S\rightarrow\ S^{-1}$.
This transformation is same as the target space
duality transformation discussed above. Notice however, that the
$SL(2,R)$ transformations discussed  here is in fact a subgroup of the
global $O(6,6)$ alluded above. The invariance of the action  under
the afore mentioned $SL(2,R)$ is due to the specific decomposition of the
background fields adopted in (12). u

We would like to point out that the $R^1\times S^1\times S^2$ wormhole
derived from (7) can be considered from another compactification mechanism.
We could consider an effective action for a closed bosonic string in critical
dimensions such that fourteen of its `internal' dimensions are flat.
Subsequently, the remaining eight internal dimensions are decomposed
into four blocks analogous to eqn. (12). Thus we could obtain four
dimensional $R^1\times S^1\times S^2$ wormholes with nontrivial $B$
and $G$ backgrounds [14].

The following remarks are in order at this stage: It is now
recognised that string theory naturally incorporates gravity and is
expected to address and resolve deep questions in quantum gravity [15].
As noted earlier the wormholes might play important role in quantum
theory of gravity. We have shown that dimensionally reduced string
effective action admits wormhole solutions and we have demonstrated
how to generate new solutions. In the recent past new cosmological
[16], and black hole [17] solutions have been obtained through
judicious choice of $O(d,d)$ transformations. We have employed only a
special class of $O(d,d)$ transformations to obtain new solutions as
illustrative examples. It is quite evident that full $O(d,d)$ as well
as $SL(2,R)$ transformations will unravel a more rich class of
solutions in addition to ones presented here.

\noindent {\bf Acknowledgements:} We are thankful to S. Kar and A. Kumar for
useful discussions.
\def \np {{\it Nucl. Phys. }}
\def \pl {{\it Phys. Lett. }}
\def \prl {{\it Phys. Rev. Lett. }}
\def \pr {{\it Phys. Rev. }}
\def \mpl {{\it Mod. Phys. Lett. }}

\noindent {\bf References:}
\item {[1]} S. W. Hawking, \pl {\bf 195B} (1987) 337; \pr {\bf D37}
(1988) 904;
G. V. Lavrelashvili, V. Rubakov and P. G. Tinyakov, {\it JETP Lett.}
{\bf 46} (1987) 167; S. B. Giddings and A. Strominger, \np {\bf B306}
(1988) 890; K. Lee \prl {\bf 61} (1988) 263.

\item {[2]} S. W. Hawking, D. N. Page and C. N. Pope, \np {\bf B170} (1980)
283; S. W. Hawking, {\it Commun. Math. Phys.} {\bf 87} (1982) 395;
S. W. Hawking, \np {\bf B244} (1984) 135.

\item {[3]} S. Coleman, \np {\bf B307} (1988) 867;
S. B. Giddings and A. Strominger, \np {\bf B307} (1988) 854.

\item {[4]} S. Coleman, \np {\bf B310} (1988) 643;

\item {[5]} J. Preskill, \np {\bf B323} (1989) 141;
I. Klebanov, L. Susskind and T. Banks, \np {\bf B317} (1989) 665;
W. Fishler, I. Klebanov, J. Polchinski and L. Susskind,
\np {\bf B327} (1989) 157; W. Fishler and L. Susskind,
\pl {\bf 217B} (1989) 48; G. W. Gibbons and S. W. Hawking,
\prl {\bf 69} (1992) 1719; {\it Commun. Math. Phys.} {\bf 148} (1982) 345;
B. Grienstein,  \np {\bf B321} (1989) 439.

\item {[6]} B. Grienstein and J. Maharana,  \np {\bf B333} (1990) 160;
S. W. Hawking, \np {\bf B335} (1990) 155;
B. Grienstein and J. Maharana and D. Sudarsky, \np {\bf B345} (1990) 231.

\item {[7]}  J. Maharana and J. H. Schwarz, Caltech. preprint,
CALT-68-1790, \np (in press).

\item {[8]}  J. Scherk and J. H. Schwarz, \np {\bf B153} (1979) 61.

\item {[9]} K. A. Meissner and G. Veneziano, {\it Phys. Lett.} {\bf 267B}
(1991) 33; G. Veneziano, {\it Phys. Lett.} {\bf 265B} (1991) 287.
 K. A. Meissner and G. Veneziano, {\it Mod. Phys. Lett.} {\bf A6}
(1991) 3397.

\item {[10]}  A. K. Gupta, J. Hughes, J. Preskill and M. Wise,
\np {\bf B333} (1990) 195.

\item {[11]} B. J. Keay and R. Laflamme, \pr {\bf D40} (1989) 2118.

\item {[12]} S. B. Giddings and A. Strominger, \pl {\bf 230B} (1989) 46.

\item {[13]} A. Shapere, S. Trivedi and F. Wilczek, \mpl {\bf A6} (1991) 2677;
A. Sen, Tata preprint, TIFR/TH/92-46 (Sep. 92); J. H. Schwarz,
Caltech preprint, CALT-68-1815; J. Maharana and J. H. Schwarz (unpublished).

\item {[14]} S. P. Khastgir and J. Maharana (in preparation).

\item {[15]} J. H. Schwarz, \pl {\bf 272B} (1991) 239.

\item {[16]} M. Gasperini, J. Maharana and G. Veneziano,
\pl {\bf 272B} (1991) 167;
M. Gasperini and G.Veneziano, \pl {\bf 277B} (1991) 256.

\item {[17]} A. Sen, \pl {\bf 271B} (1991) 295;
A. Sen, \pl {\bf 274B} (1991) 34;
S. F. Hassan and A. Sen, \np {\bf B375} (1992) 103;
S. K. Kar, S. P. Khastgir and A. Kumar, \mpl {\bf A7}
(1992) 1545; A. Sen, \prl {\bf 69} (1992) 1006.

\end